# Mining Financial Statement Fraud

An Analysis of Some Experimental Issues

Jarrod West
School of Computing & Mathematics
Charles Sturt University, Australia
Albury, Australia-2640
jnwest@netspace.net.au

Maumita Bhattacharya
School of Computing & Mathematics
Charles Sturt University
Albury, Australia-2640
mbhattacharya@csu.edu.au

*Abstract*—Financial statement fraud detection is an important problem with a number of design aspects to consider. Issues such as (i) problem representation, (ii) feature selection, and (iii) choice of performance metrics all influence the perceived performance of detection algorithms. Efficient implementation of financial fraud detection methods relies on a clear understanding of these issues. In this paper we present an analysis of the three key experimental issues associated with financial statement fraud detection, critiquing the prevailing ideas and providing new understandings.

## I. Introduction And Background

Financial statements are official documents that are released by companies to document their practices over a period of time. They typically cover details such as profit expenditure and loans, as well as statements by managers, directors, and board members. A well-known example of financial statements are the 10-K reports required to be filed by large U.S. companies for the Securities and Exchange Commission every year [7], [10]. Financial statement fraud occurs when a company falsifies these documents for financial gain. As financial statements reflect a snapshot of the business's performance during a specific period they have a large impact on the company's share prices, borrowing power, and consumer confidence. By altering financial statements unscrupulous managers can improve the public appearance of the company, often leading to an unjustified increase to income. Reasons for committing this fraud may vary from pure financial gain through stock and bonuses, the disguising of managerial ineptitude, or as a genuine attempt to help the business through a difficult period [10], [13]. For our purposes we will also consider accidental fraud caused either by incompetence or inexperience, as this can have the same financial ramifications as intentional fraud and is therefore desirable to be detected.

Over the years various computational methods have been used for financial statement fraud detection and, like other similar problems, successful implementation of the detection methods depends on clear understanding of the problem domain. Though many prior researchers have investigated financial statement fraud with individual experiments there has been a lack of research into analysis of the problem domain that we aim to address here.

Kirkos et al. [3] looked at features designed to represent financial distress, high debt structure, and false growth. They used ANOVA to determine the significance of the features and tested them with three common classification methods. Hoogs et al. [1] studied the reports of individual companies, comparing features to their historical values and those of similar businesses in their industry. They then introduced a genetic algorithm which was effective at detecting fraud in those companies. Yue et al. [2] performed a review of the financial fraud detection research available at that time and found that feature selection was an important aspect that had not been given sufficient investigation by researchers. Additionally they also noted the disparity of research between specific detection methods. Ravisankar et al. [10] studied financial ratios encapsulating the concepts of liquidity, safety, profitability, and efficiency. They then compared the capabilities of several data mining methods including neural networks, genetic programming, support vector machine, group method of data handling, and logistic regression. Zhou and Kapoor [13] studied the choices that management can take when presented with financial hardship, including the environmental factors that are often present when fraud is committed. They then proposed an adaptive framework to be used as the basis for financial statement fraud detection methods.

Ngai et al. [4] looked at the distribution of prior research into various types of fraud and noticed that that there was an absence of study into visualisation and outlier methods. Glancy and Yadav [5] and Zaki and Theodoulidis [9] both looked at text present in financial statements. The former showed that the text frequently contained hidden indicators of deception that could be uncovered after processing. The latter used text from litigation cases, utilising the formal structure of the documents to reduce the complexity of the problem resulting from the qualitative nature of the data. Huang [11] made use of typical financial ratios as well as corporate governance indicators to provide input to a support vector machine. He used various forms of z-score to assert the separation between fraudulent and legitimate companies as well as the t-test to determine the effectiveness of the selected features. West et al.

[7] studied financial statement fraud as part of a broad review and concluded that there is a requirement for further research into the specific problem domains of each fraud type and the comparative performance of detection methods.

As stated before, this paper aims at providing better insight into experimental issues associated with financial statement fraud detection, which could form the basis for effective implementation of fraud detection methods. For our analysis, we have considered the following key experimental issues: (i) *problem representation*; (ii) *feature selection* and (iii) *performance metrics*.

The rest of the paper is organised as follows: Section II focuses on various problem representation models suitable for financial statement fraud detection problems; Section III presents an analysis on feature selection and Section IV details a selection of performance metrics that can be used in this problem domain. Finally Section V provides some concluding remarks.

## II. Problem Representation

To be capable of solving a complex problem like financial statement fraud it is important to first obtain a complete understanding of the problem domain. Problem representation will have significant ramifications on the outcome of detection methods and therefore requires careful consideration. This section covers the various models that can be used for financial statement fraud detection.

### A. Regression analysis

Regression is a statistical method that aims to expose relationships between a dependent variable and a set of independent variables [4]. Regression analysis is a suitable model because of its proven experience with fraud detection and similar problems like network intrusion, as well the ease with which it is implemented and understood. However, regression may be less refined than other models and therefore give inferior results [7].

### B. Classification

Classification is a data mining method that aims to separate a list of unknown samples into one of several discrete classes [4]. Binary classification is a simplified case in which there exists only two possible categories (such as fraudulent and non-fraudulent). Classification problems are suitable as the solutions are powerful, capable of solving non-linear problems, and typically require very little knowledge of the specific problem domain. However, they can also be computationally inefficient and susceptible to overtraining [7].

### C. Visualisation

Visualisation refers to any method that results in the presentation of data into clear and understandable format for the purpose of being manually observed by a human operator. Like other models, visualisation seeks to determine underlying relationships or concepts within the problem, but instead of acting on that information directly it processes it into readable patterns for users [4], [6]. The major benefit of any type of visualisation is that it can simplify multi-dimensional results down into a format that is understood by humans. Downsides include a difficulty in correctly interpreting complex results and the lack of complete automation [8].

### D. Clustering

Similar to classification, clustering is a method that splits samples into distinct, related groups that have no affiliation to other categories [4]. A clustering model makes use of a measure of similarity to assign input samples to clusters within a dimensional space: samples which are calculated to have a high similarity are naturally grouped together into the same cluster. Clustering offers an ease of implementation and comprehension but can be computationally inefficient and have difficulties with problems with noisy data, such as financial statement fraud.

### E. Rule-based problem

Association rules offer a simple form of classification based on established mathematical logic statements. A model is created that takes a set of attributes and forms a prediction on the outcome, such as the probability that a particular sample is fraudulent or not. Association rules are suitable as they are easy to understand and mathematically sound, but they can also be vulnerable to overtraining.

## III. Feature Selection

Regardless of the detection model chosen, feature selection is an integral part of solving any problem. Each method relies on processing large quantities of data to detect obscured relationships and meanings, and therefore the variables selected for inclusion must be a good representation of the data as a whole. For financial statement fraud any part of the financial document is a feature that may be used in detection algorithms, including all kinds of numerical, categorical, and textual data.

The aim of feature selection is to improve both the actual and computational performance of the solution, as well as providing a better understanding of the problem. To this end, algorithms are used to rank or choose which features are the most applicable to the current task. Feature ranking algorithms make use of an evaluation method to assign a rating to individual features based on attributes such as consistency, accuracy, and content, and choose a subset of these based on that ranking. When used correctly this subset should have comparable ability to the full set while being significantly smaller. Feature selection is a required part of data preprocessing, however it may also be used as part of the data mining algorithm itself [8].

Considering the number of features available, feature selection can be a severely time consuming task: assuming an initial set of $n$ features, an exhaustive search of the best possible combination to use would require $2^n$ comparisons. For this reason selection algorithms typically use heuristic measures to reduce the number of evaluations that are required. This step requires careful consideration, as overzealous pruning can remove beneficial features and result in reduced accuracy for any solution. After the selection is completed, a further step known as feature extraction is undertaken to transform the problem's input set into a feature set that is ready for the data mining algorithm to be applied [8].

## A. Commonly used features for financial statement fraud

The most common features chosen for financial fraud detection are numerical, and based around the company's earnings, assets, and expenditures. A list of the typical variables used is given in Table I [1], [10].

TABLE I.  COMMON FEATURES USED FOR FINANCIAL STATEMENT FRAUD DETECTION

| Feature | Description | Representation |
|---|---|---|
| Debt | Amount of money owed in the form of loans or bonds | D |
| Assets | The total value of all items held by the company such as inventory, property, or cash | A |
| Gross and net profit | The amount of financial gain made in a given time period, either absolute or relative to expenses | $P_G$, $P_N$ |
| Cash and deposits | The gross amount of funds available to the company | C |
| Inventory | The value of current available product, or the saleable portion of assets | I |
| Expenses | Total amount of costs incurred by the company in a given time period | X |
| Equity | The total amount of funds contributed by shareholders | E |
| Sales | The raw amount of product that has been sold by the company to their customers | S |

The quantitative nature of these features makes them easily adaptable to many detection algorithms. However, given that the details of a particular company can vary significantly, using absolute values for these numerical fields is not ideal. A suitable solution is to utilise financial ratios, considering the size of a particular value relative to another. Some examples of common financial ratios that are used are given in Table II [1], [3], [10].

Note that some of these ratios involve an element of time, focussing on a specific period or values relative to previous years. Time is a highly beneficial measure to detect changes within a company that could indicate the beginnings of fraudulent behaviour [1].

## B. Methodology for choosing features for financial statement fraud

A common approach to financial statement fraud feature selection is the use of the z-score, a performance metric that many researchers have used to determine whether there is significant difference between features [1], [3], [10]. Others make use of the analysis of variance method (ANOVA) to accomplish the same task [3]. In addition to these statistical methods many researchers choose their features based on existing fraud detection studies or domain based reasoning.

TABLE II.  COMMON FINANCIAL RATIOS USED FOR FINANCIAL STATEMENT FRAUD DETECTION

| Feature | Description | Representation |
|---|---|---|
| Debt to equity | Ratio of debt over equity | D/E |
| Sales to total assets | Ratio of sales over assets | S/A |
| Earnings before interest and taxes | Sales minus expenses, also known as EBIT | S-X |
| Working capital | Current assets minus current liabilities | W |
| Total debt to total assets | Ratio of debt over assets | D/A |
| Net profit to total assets | Ratio of net profit over assets | $P_N$/A |
| Working capital to total assets | Ratio of working capital over assets | W/A |
| Gross profit to total assets | Ratio of gross profit over assets | $P_G$/A |
| Cash to total assets | Ratio of cash over assets | C/A |
| Total assets to total assets of previous year | Ratio of current assets over the previous year's assets | A/$A_P$ |
| Long term debt to total assets | Ratio of debt planned for more than one year over assets | $D_L$/A |
| Net profit to gross profit | Ratio of net profit over gross profit | $P_N$/$P_G$ |
| Total growth | Difference in sales from previous year | (S-$S_P$)/$S_P$ |

As explained previously there are a variety of reasons for committing financial statement fraud, from guiding the company through difficult times to disguising managerial incompetence. In practice, financial statement fraud typically takes one or more of the following forms: failing to report or undervaluing revenues, overstating the value of existing assets, understating the value of expenses, or outright theft or embezzlement of assets [13].

With this in mind it is logical to reason that the relevant financial ratios will focus heavily on assets, sales, and expenses. Additionally there are indicators within a company that it may turn to fraud, such as financial distress, pressure, and a requirement for growth [3]. This focus on public-perception of the company encourages analysis of the features often reported in the media, such as total growth and gross and net profits.

Debt is an attribute that has been used extensively in prior financial fraud detection research, as there is strong evidence correlating companies with high debt structures and fraudulent behaviour [3], [10]. A high debt structure is encompassed with attributes such as debt to equity and debt to assets. Imbalances between incoming and outgoing cash flow can be identified by comparing sales to total assets, net profit to gross profit, and gross profit to total assets. Underrepresented expenses are covered with earnings before interest and taxes, and working capital to total assets.

The importance of time as a component in financial statement fraud has been observed by several researchers. Introducing ratios based on prior performance can identify suspicious changes within financial statements [1]. Total assets to total assets of previous year and total growth are both capable of demonstrating time-based financial statement differences.

Despite the readily available numerical data some researchers focus on features derived from the subjective information included within financial statements. Many governing bodies require that companies include various forms of plaintext with their statements, including managerial announcements and notices to shareholders, and these can contain indicators of fraudulent behaviour. Selecting text as a feature offers an additional level of complexity but can provide crucial additional evidence to the solution. When considering the selection of text as a feature a preprocessing step is often applied to transform it into a quantitative form which can then be used directly [5], [9].

*C. Potential issues with feature selection*

To create the training set that is required for many fraud detection solutions a selection of legitimate companies must be chosen to serve as the control for non-fraudulent behaviour. To reduce unintentional interference from introduced variables companies of a similar size and function are typically chosen for this task. This results in a far smaller sample size than the actual problem domain, with researchers often using matched pairs of one legitimate company to one fraudulent [1].

As well as this, the controls are selected based on the lack of evidence of fraud: there is no guarantee that these are legitimate companies or statements, and using fraudulent values here would greatly diminish the ability of the data mining solutions. Further, much of the above reasoning for determining features begins with the assumption that managers are intentionally committing fraud, for the benefit of the company or their own personal gain. This doesn't account for the case where the fraud is accidental, due to either mismanagement or incompetence, which is an additional scenario that would be useful to detect.

## IV. PERFORMANCE METRICS

Measuring the success of data mining algorithms is an important step in determining their suitability at solving their respective problem. This is especially true for a problem such as financial statement fraud, where minor improvements in performance can lead to large economic benefits. Performance can be measured in many different ways: *absolute ability*, *performance relative to other factors*, *visual mediums*, *probability of success*, and more. In this section we will first define a range of performance metrics that is suitable for financial statement fraud detection, followed by an analysis of the relevant issues surrounding each one.

*A. Binary classification metrics*

As financial statement fraud is a binary problem there are only two possible outcomes, which we measure as:

- Positive (P). The number of samples that are actually positive.
- Negative (N). The number of samples that are actually negative.

Some additional derived ratios are used as the basis of the more complex metrics listed below [12].

- True positive (TP). The number of samples that are classified as positive and are actually positive.
- True negative (TN). The number of samples that are classified as negative and are actually negative.
- False positive (FP). The number of samples that are classified as positive but are actually negative. Also known as Type I error.
- False negative (FN). The number of samples that are classified as negative but are actually positive. Also known as Type II error.

*1) Accuracy:* The number of samples correctly classified as a percentage of the total samples.

$$ACC = (TP+TN)/(P+N) \quad (1)$$

*2) Sensitivity:* The number of positive samples correctly classified as a percentage of the total positive samples. Also known as recall, true positive hit rate, or hit rate.

$$SENS = TP/P \quad (2)$$

*3) Specificity:* The number of negative samples correctly classified as a percentage of the total negative samples.

$$SPEC = TN/N \quad (3)$$

*4) Precision:* The number of positive samples correctly classified as a percentage of total samples correctly classified.

$$PREC = TP/(TP+FP) \quad (4)$$

*5) False positive rate:* The inverse of the true positive rate, given as 1-specificity, which can be simplified to:

$$FPR = FP/N \quad (5)$$

*6) F-measure:* Also known as F-score or F, the F-measure is the harmonic mean of precision and recall (sensitivity) [6], [12]. It is given by:

$$F = 2/(1/PREC+1/SENS) \quad (6)$$

Which can be simplified to:

$$F = (2 \times PREC \times SENS)/(PREC+SENS) \quad (7)$$

*7) Fβ:* A form of F-measure that applies a weighting of *β* to the precision and recall, where *β* is a positive, real number [6].

$$F_\beta = ((1+\beta^2) \times PREC \times SENS)/(\beta^2 \times PREC + SENS) \quad (8)$$

*8) Cost minimisation:* Cost minimisation is a study of the effectiveness of an algorithm using the misclassification costs for each type of error. Minimising this value provides a valuable measure of the real-world suitability of a solution. The simplest form uses the false negative and false positive rates, weighted by their costs:

$$C = FPR \times C_{FP} + FNR \times C_{FN} \quad (9)$$

Where $C_{FP}$ and $C_{FN}$ are the costs incurred with a false positive and false negative respectively.

B. *Statistical metrics*

*1) ANOVA:* Short for analysis of variance, ANOVA is a method of analysing regression and other data mining solutions by focussing on the variance between observed results and expected values. ANOVA is a blanket term for many types of variance metrics [8].

*2) Z-score:* Used to measure the rate of change in a variable, either independently, with respect to its historical values, or against a similar variable.

$$Z = (X-\mu)/\sigma \quad (10)$$

Where $X$ is the raw value of the variable and $\mu$ and $\sigma$ are the mean and standard deviation of the variable respectively [11].

*3) Z-within score:* The ratio of change in a variable from its historical values:

$$Z_w = (X - \mu_h)/\sigma_h \quad (11)$$

*4) Z-between score:* The ratio of change in a variable from similar values:

$$Z_b = (X - \mu_c)/\sigma_c \quad (12)$$

*5) Sum of squared errors:* A measurement of the difference between two sets of values, squared to separate out distinct clusters of values.

$$E = \sum_{i=1}(y_{Ei} - y_{Ai})^2 \quad (13)$$

Where $y_{Ei}$ is a known, expected result for input $i$, and $y_{Ai}$ is the result obtained by the solution. This makes $y_{Ei} - y_{Ai}$ the given measure of error.

C. *Association rule metrics*

*1) Support:* An itemset is a group of items that commonly occur together across the problem space. Given this, the support for a rule is a measure of the percentage of samples that contain the itemset:

$$S_I = (\sum X_{Ij})/N \quad (14)$$

Where $S_I$ is the support for itemset $I$, $X_{Ij}$ is a sample that contains the itemset $I$, and $N$ is the total number of samples [8].

*2) Confidence:* Given a rule $X \rightarrow Y$, where $X$ and $Y$ are subsets of an itemset $I$ and $X \cap Y = \emptyset$, the rule has a confidence equal to the proportion of samples that contain both $X$ and $Y$ over the total that contain $X$. Expressed in terms of the support, this becomes:

$$C_{X \rightarrow Y} = S_{X \cup Y}/S_X \quad (15)$$

Where $C_{X \rightarrow Y}$ is the confidence of the rule, $S_X$ is the support of itemset $X$, and $S_{X \cup Y}$ is the support of an itemset formed by the union of itemsets $X$ and $Y$ [8].

*3) Lift:* A correlation measure used to determine whether an association rule is useful to the problem, lift is given by:

$$L_{X \rightarrow Y} = C_{X \rightarrow Y}/S_Y \quad (16)$$

and measures the probability that a change to $X$ will result in a corresponding change to $Y$. A result of greater than one indicates a positive correlation, implying that the occurrence of one results in an occurrence of the other, and a result less than 1 indicates a negative correlation, suggesting that the occurrence of one results in an absence of the other. A value of exactly 1 indicates that both variables are completely independent of each other [6].

*4) Conviction:* Given the same type of rule as given for confidence above, the conviction is a measure of the inaccuracy of the rule, or the chance of $X$ occurring without $Y$:

$$CV_{X \rightarrow Y} = (1-S_Y)/(1-C_{X \rightarrow Y}) \quad (17)$$

Where $CV_{X \rightarrow Y}$ is the conviction of the rule, $S_Y$ is the support of itemset $Y$, and $C_{X \rightarrow Y}$ is the confidence of the rule [8].

D. *Clustering metrics*

*1) Hopkins statistic.* The Hopkins statistic is a measure of the probability that a variable is randomly distributed within a space. It can be used to determine whether a dataset contains significant clusters, and is given by:

$$H = (\sum_{i=1} y_i)/(\sum_{i=1} x_i + \sum_{i=1} y_i) \quad (18)$$

Where $x_i$ is the distance to the nearest neighbour for point $i$ in a uniformly distributed version of the dataspace, and $y_i$ is the distance to the nearest neighbour for point $i$ in the sampled dataspace. If the samples were distributed evenly $y_i$ would approach $x_i$ and $H \approx 0.5$. However, the further the variable is from uniformly distributed the smaller the value of $y_i$, resulting in $H > 0.5$.

E. *Visual metrics*

*1) ROC curve.* Standing for receiver operating characteristic, an ROC curve is a two dimensional graph modelling the true positive rate on the Y-axis against the false positive rate on the X-axis [6], [12]. ROC curves provide an easily interpreted visualisation of the success of a binary classification method, as well as identifying a few key areas of interest.

The point at (0,1) represents perfect accuracy with no false positives. The origin (0,0) reflects no true positives and no false positives, indicating that no samples were classified positive at all. The point at (1,1) shows the opposite with all samples classified as negative. With these features an ROC graph can be used to help understand the reason behind a classification method's success [12].

*2) AUC.* The area under an ROC curve, given as a value between 0 and 1. By assessing both true positive and false positive rates it gives a single-value measure of the accuracy of a model [12].

*F. Analysis and Observations*

There is a large variety of *classification metrics* which offer several different approaches to determining solution performance. Accuracy is a very useful all-round method that works well in many cases, but it suffers when there is a large disparity between positive and negative samples. For financial statement fraud detection sensitivity and specificity may be more useful. Cost benefit analysis would be ideal but would require a highly accurate estimate into the cost of false positives and false negatives, a non-trivial task.

*Statistical metrics* are comprised of a number of dependable and well-understood error and rate of change measurements. Additionally many statistical methods are applicable for use with other aspects of data mining, including feature selection and the solution algorithm itself. Error calculations like the sum of squared errors can be used to separate samples for classification and clustering, and ANOVA and the Z-score can determine whether two variables are significantly different.

*Association rule metrics* offer a number of performance measurements for individual rules. Support, confidence, and conviction all provide measurements to determine how accurate a rule is, and lift indicates whether a rule is useful to a specific problem domain. However, there are no standard metrics that measure the overall effectiveness of an association rule solution, so a combination with some other measurement is likely required for financial fraud detection.

There is a single key metric for testing the effectiveness of *clustering* algorithms, and the Hopkins statistic only measures the significance of the results, not their overall performance. Clustering results can often be represented graphically and viewed by a human operator but will be unsuitable for automated testing.

*Visual metrics* provide a way for auditors to quickly and easily comprehend the detection results. Despite its straightforward appearance the ROC curve provides extensive information on binary classification results, including all the points of interest for true and false positive rates. Additionally the area under the curve adds a qualitative facet to visual results.

V. CONCLUSION

Financial statement fraud is primarily a significant problem within our modern society, and its successful detection requires a detailed understanding of the problem domain and the issues that surround it. This paper analysed the key experimental issues involved with financial statement fraud detection, namely problem representation, feature selection, and performance metrics, identifying potential problems and offering recommendations on how to avoid them. Our purpose is to provide a reference for future researchers and practitioners to utilise when undertaking their own experiments.

There are still aspects of financial statement fraud detection that would warrant further investigation. Additional research could explore the issues in greater depth, particularly with a focus on specific detection methods. Also, future researchers may wish to concentrate on additional comparisons between the various performance metrics using controlled experiments.